\documentclass[twocolumn,showpacs,prl,amsmath,amssymb]{revtex4}
\usepackage{graphicx}% Include figure files
\usepackage{epsfig}
\usepackage{dcolumn}% Align table columns on decimal point
\usepackage{amsmath}
\usepackage{bm}% bold math

\begin{document}

\title[Short Title]{Exciton energy transfer in nanotube bundles}
\author{P. H. Tan, A. G. Rozhin, T. Hasan, P. Hu, V. Scardaci,W. I. Milne}
\author{A. C. Ferrari}
\email{acf26@eng.cam.ac.uk} \affiliation{Department of Engineering,
University of Cambridge, Cambridge CB3 0FA, UK}

\begin{abstract}
Photoluminescence is commonly used to identify the electronic
structure of individual nanotubes. But, nanotubes naturally occur in
bundles. Thus, we investigate photoluminescence of nanotube bundles.
We show that their complex spectra are simply explained by exciton
energy transfer between adjacent tubes, whereby excitation of large
gap tubes induces emission from smaller gap ones via
F$\ddot{o}$rster interaction between excitons. The consequent
relaxation rate is faster than non-radiative recombination, leading
to enhanced photoluminescence of acceptor tubes. This fingerprints
bundles with different compositions and opens opportunities to
optimize them for opto-electronics.
\end{abstract}

\date{January 15, 2007}

 %PACS, the Physics and Astronomy Classification Scheme.

\maketitle

Carbon nanotubes are rolled graphene sheets~\cite{rbook}. One
dimensional quantum confinement makes their band structure
fundamentally different from graphene, with sub-bands and
singularities in the density of states~\cite{rbook}. These are fully
determined by their chiral indexes (n,m), which specify how graphene
is folded into each nanotube. Thus, measuring the optical
transitions allows in principle to determine the chiral indexes, and
fully identify a nanotube sample. For this reason, a massive effort
was put to measure photoluminescence in nanotubes since their
discovery. However, it took more than ten years to unambiguously
detect and identify photoluminescence emission from single wall
nanotubes (SWNT)\cite{Connell,Bachilo1}. Indeed, SWNT naturally
occur in bundles, due to Van-der-Waals interactions~\cite{Hertel},
but a significant intensity was only measured when efficient
de-bundling was achieved\cite{Connell,Bachilo1}. This paved the way
to the interpretation of their complex absorption and emission
spectra\cite{Connell,Bachilo1}. The discrepancy between
single-particle theory and experiments, pointed to the major role of
electron-electron and electron-hole interactions in shaping their
band-structure\cite{Wang1,Maultzsch}. The exciton binding energies
were recently measured\cite{Wang1,Maultzsch}. These are very large,
from tens meV to 1 eV, depending on diameter, chirality, and
dielectric screening~\cite{Wang1,Maultzsch,Perebeinos2}. Thus,
excitons dominate even at room temperature.

The investigation of the optical properties of nanotubes is now a
most pursued research
area\cite{Connell,Bachilo1,Misewich,Chen,Perebeinos2,Wang1,Maultzsch,Ma},
however this still focuses on individual tubes, in contrast with
their natural occurrence in bundles. Furthermore, the luminescence
quantum yield of individual SWNTs is very low and this hinders their
applications in optoelectronics~\cite{Bachilo1,Misewich,Chen}. Here
we present a thorough investigation of absorption and emission
spectra of nanotube bundles. We show that their apparently complex
spectra can be simply interpreted considering exciton energy
transfer between tubes. This is a well known phenomenon in
biological systems, conjugated polymers, quantum wires, dots, and
other low-dimensional systems\cite{Becker,Biju,Forster,Rizzo,Kagan},
which we now clearly identify in nanotubes. We find that energy
transfer is a major non-radiative relaxation channel for large gap
tubes, strongly enhancing the photoluminescence of the acceptor
tubes. Thus, contrary to what usually assumed, nanotube bundles
could be ideal for high yield optoelectronics applications, far
surpassing the poor performance of individual
tubes~\cite{Misewich,Chen}. Furthermore, energy transfer
fingerprints bundles with different nanotube concentrations, finally
offering a quantitative means to monitor the composition of bundles
in solution, a key requirement for research and
applications\cite{Connell,Bachilo1}.

We measure absorption on CoMoCAT SWNT(South West
Nanotechnologies)~\cite{Kitiyanan} suspensions in D$_2$O with sodium
dodecylbenzene sulfonate (SDBS) surfactant\cite{Connell}, using a
Perkin-Elmer 950 spectrometer. A JY Fluorolog-3 is used for
Photoluminescence Excitation (PLE).
\begin{figure}
\centerline{\includegraphics[width=80mm]{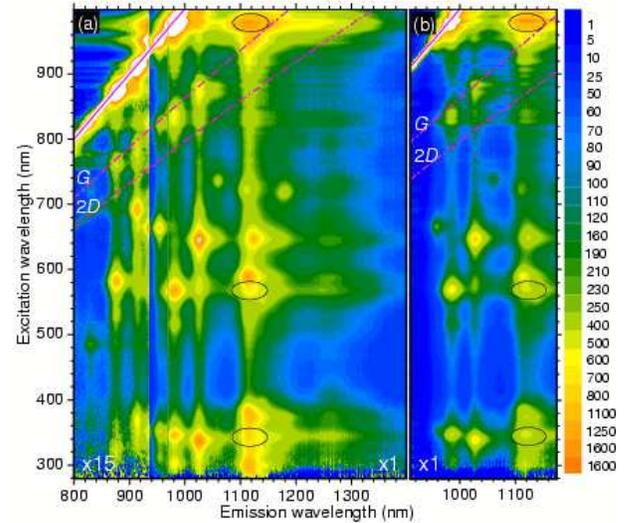}}
\caption{PLE map for (a) as-prepared suspensions and (b) after two
months. Solid lines at upper left corners represent resonances
with same excitation and recombination energies. The dashed-dotted
lines are associated with G and 2D sidebands. Circles mark
emission from (8, 4), (7, 6) and (9, 4) SWNTs, with excitation
matching $eh_{11}$, $eh_{22}$,$eh_{33}$ of (6, 5).} \label{fig:1}
\end{figure}

Figure 1a plots PLE maps from the as-prepared solution. Each spot
can be labeled as ($\lambda_{ex}$, $\lambda_{em}$), where
$\lambda_{ex}$, $\lambda_{em}$ are, respectively, the excitation and
emission wavelengths. Several high intensity peaks in Fig.1a are
exciton-exciton resonances~\cite{Bachilo1,Bachilo2}. In this case
$\lambda_{ex}$ corresponds to the energy of the excitonic states
$eh_{ii}$ associated with the $i$th electronic interband transitions
$E_{ii}$ ($i$=1, 2, 3, 4) in the single particle picture
~\cite{Bachilo1,Bachilo2}, while $\lambda_{em}$ is the emission
energy of the lowest exciton transition $eh_{11}$. Other spots in
Fig.1a are related to exciton-phonon
sidebands~\cite{Perebeinos1,Plentz,Miyauchi}. The spectral features
in Fig.1a are summarized in Fig.2. We identify 16 different SWNT
species in the range 800nm-1300nm, and assign their chiral indexes
in Fig. 2, as for Ref.~\onlinecite{Bachilo1}. The phonon sidebands
for the $eh_{11}$ and $eh_{22}$ excitons are shown in Fig. 2 with
open circles and diamonds. The $eh_{ii}$ wavelengths of most SWNT
here are 3-10 nm larger than Ref. \onlinecite{Bachilo1}. This
red-shift is expected in the presence of
bundling~\cite{Reich3,Wang}. Fig.2 has some interesting features
compared with previous data on isolated SWNT
suspensions~\cite{Bachilo1,Bachilo2}: 1) the spectral profile of
exciton resonances significantly elongates in the horizontal and
vertical directions; 2) new  peaks appear, such as, e.g., (645nm,
1265nm) and (568nm, 1250nm), with intensity much stronger than the
($eh_{22}$, $eh_{11}$) peaks of (10, 5), (8, 7) and (9, 5) SWNTs; 3)
a strong broad band near (980nm, 1118nm) is observed.

\begin{figure}
\centerline{\includegraphics[width=80mm]{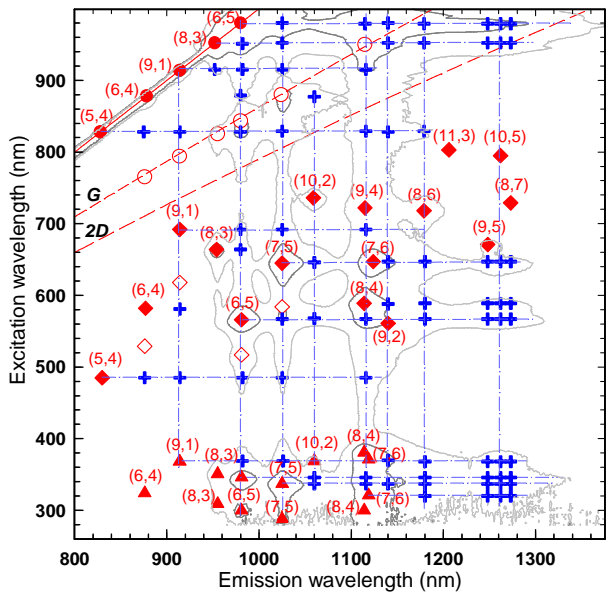}}
\caption{Main features in the PLE spectra of as-prepared
suspension. Solid circles, diamonds and triangles represent
$eh_{11}$ emission of SWNTs for which excitation matches their
$eh_{11}$, $eh_{22}$, $eh_{33}$ and $eh_{44}$ transitions. Each
peak is labeled with the chiral index of the corresponding SWNT.
Open circles and diamonds are phonon sidebands. Solid crosses are
assigned to EET between s-SWNTs. Gray contour patterns comprise
both exciton-related resonances and EET spectral features.}
\label{fig:2}
\end{figure}

To clarify the origin of these bands, we check PLE from the same
solution after two months (Fig.~1b). Figs.~1a,b have similar
features. However, the $eh_{11}$ emission wavelengths of most SWNTs
in Fig.~1b redshift$\sim$3-5 nm relative to Fig.~1a. This suggests
further aggregation into bigger bundles. Also, almost all spots
decrease in intensity. But a careful examination of Figs.~1a,b,
shows that the (980nm, 1118nm) peak becomes stronger after two
months. Also, two peaks near (568nm, 1118nm) and (346nm, 1118nm),
indicated by circles in Figs.~1a,b, are identified more clearly, due
to the lower intensity of the ($eh_{ii}$, $eh_{11}$)($i$=2, 3, 4)
bands of (8, 4) and (7, 6) tubes, which shadowed them in the
pristine solution. Notably, these three peaks do not correspond to
any of the known exciton-exciton resonances of SWNTs in this
spectral range~\cite{Bachilo1,Bachilo2}. The (980nm, 1118nm) peak is
not assigned to a $D$ phonon sideband of (8, 4), (7, 6) or (9, 4)
tubes, due to the lack of such sideband in previous investigations
of these and other tubes~\cite{Perebeinos1,Plentz,Htoon,Chou}.
Indeed, the excitation energies of the (980nm, 1118nm), (568nm,
1118nm) and (346nm, 1118nm) bands match, respectively, the
$eh_{11}$, $eh_{22}$ and $eh_{33}$ transitions of (6, 5)
tubes\cite{Bachilo1}, whereas their emission around 1118 nm is
consistent with (8, 4), (7, 6), (9, 4) $eh_{11}$. Thus, resonant
excitation of large gap donors tubes induces emission from smaller
gap acceptors. This implies energy transfer between SWNT in bundles.
Due to the large exciton binding energies\cite{Wang1,
Maultzsch,Perebeinos2}, this happens mainly via excitons, not via
intertube electron or hole migration~\cite{Torrens}.

A thorough examination of all peaks in Figs.~1a,b, allows us to
identify several other exciton energy transfer (EET) features (solid
crosses in Fig.~2). The peaks not attributable to known
exciton-exciton resonances along each horizontal dashed-dotted line
in Fig.2 are assigned to $eh_{ii}$ excitation of donor tubes,
inducing $eh_{11}$ emission from acceptors. Vice-versa, the crosses
along each vertical dashed-dotted line are $eh_{11}$ emission of an
acceptor, following EET from $eh_{ii}$ excitation of donors.
Broad/elongated patterns of Fig.1, shown by grey contours in Fig.2,
contain overlapping peaks from tubes with similar excitation or
emission energies. Size, concentration and distribution of nanotube
species within a bundle will determine the EET-induced intensities.
The higher the concentration of semiconducting tubes, the higher the
probability of them being adjacent, the higher the chance of
EET-induced emission. Thus, the strongest peaks will appear around
$eh_{ii}$ transitions of semiconducting tubes with highest
concentration, such as (6, 5), (7, 5), (8, 4) in our CoMoCAT
solutions~\cite{Kitiyanan,Bachilo2}.

\begin{figure}
\centerline{\includegraphics[width=72mm]{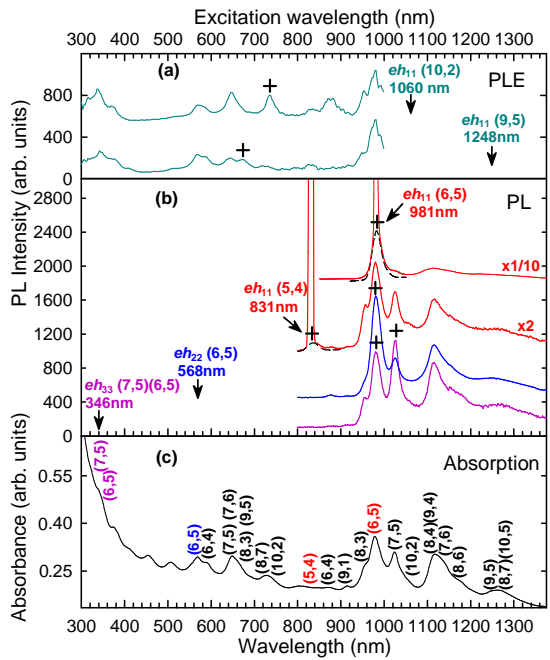}}
\caption{(a) PLE spectra, (b) emission spectra and (c) absorption
spectrum. Arrows indicate detection wavelengths in PLE and
excitation in emission. Exciton resonances and associated (n, m)
are also shown. Crosses in (a,b) mark exciton-exciton resonances
in emission and excitation. Dashed lines in (b) are fits of the
($eh_{11}$, $eh_{11}$) resonances. These fits are done after
subtracting the fitted Rayleigh peaks of the SWNT solution from
that of a D$_2$O/SDBS solution without SWNTs.} \label{fig:3}
\end{figure}

Fig.3 compares the absorption of the as-prepared solution with its
photoluminescence emission and excitation spectra. The ($eh_{ii}$,
$eh_{11}$) peaks are marked by crosses. We assign most of the other
bands to energy transfer from donor to acceptors within bundles. The
($eh_{11}$, $eh_{11}$) peak is the strongest amongst all possible
($eh_{ii}$, $eh_{11}$) for a given (n, m), as, e.g., in the (6, 5)
tube of Fig.~3b. This is because $eh_{11}$ excitons have higher
density of states than $eh_{22}$, $eh_{33}$~\cite{Spataru0}. Thus
more photons are absorbed by $eh_{11}$ states. Then, as shown in
Fig. 3a, the $eh_{11}$ excitation of large gap donors is a more
efficient way to enhance emission of smaller gap acceptors than the
direct $eh_{22}$ and $eh_{33}$ excitation of the acceptors, despite
the low photoluminescence quantum efficiency of individual
tubes~\cite{Connell}.

We can estimate the exciton energy transfer efficiency in bundles as
follows. Let us consider the exciton relaxation of two adjacent
tubes with different gaps, following the resonant $eh_{11}$
excitation of the larger gap tube. The rate equations of the
donor-acceptor system are:
\begin{equation}
\partial n_D/\partial t=G_{pe}-n_D(1/\tau_{nrD}+1/\tau_{rD})-n_D/\tau_{DA}
\label{first}
\end{equation}
\begin{equation}
\partial n_A/\partial t=n_D/\tau_{DA}-n_A(1/\tau_{nrA}+1/\tau_{rA})
\label{second}
\end{equation}
where $\tau_{DA}$ is the energy transfer lifetime between donor and
acceptor, $n_D$ is the population of excitons in the donor and $n_A$
in the acceptor, $\tau_{nrD}$, $\tau_{rD}$, $\tau_{nrA}$ and
$\tau_{rA}$ are the radiative (r) and non-radiative (nr) lifetimes,
$G_{pe}$ the exciton density in the donor created by
photo-excitation. An estimation of the EET efficiency is the ratio
of acceptor $eh_{11}$ emission intensity ($I_A$=$n_A/\tau_{rA}$) to
that of the donor ($I_D$=$n_D/\tau_{rD}$). Then, deriving $n_A/n_D$
from Eqs. (1,2) at steady state, we get:
\begin{eqnarray}
I_A/I_D=\frac{1/\tau_{DA}}{1/\tau_{rA}+1/\tau_{nrA}}\frac{\tau_{rD}}{\tau_{rA}}
\label{third}
\end{eqnarray}
The $eh_{11}$ radiative lifetime is reported $\sim$20-180ps,
dependent on temperature~\cite{Hagen}. For tube diameters
$\sim$0.75-0.95 nm, it is about $\sim$20-30 ps at room
temperature\cite{Hirori}. This is much shorter than the theoretical
radiative lifetime ($\sim$10 ns)\cite{Spataru}. Thus, the observed
lifetimes are determined by non-radiative recombination. Eq. (3) can
then be simplified as $I_A/I_D\approx\tau_{nrA}/\tau_{DA}$.

We measure a very high $I_A/I_D$ in bundles. E.g., under $eh_{11}$
excitation of the (5, 4) tubes in Fig.~3b, the ratio of
photoluminescence intensity of all acceptor tubes with emission
above 900 nm [such as (6, 5), (7, 5), (8, 4),(7,6)] to that
at$\sim$831 nm of the (5, 4) donor tubes is at least$\sim$75. This
indicates that most resonantly-excited (5, 4) $eh_{11}$ excitons
transfer their energy to the acceptors, rather than recombine. Thus,
in bundles exciton relaxation is comparable or even faster than
non-radiative recombination. This fast relaxation suppresses
emission from donors. But it significantly increases the acceptors
luminescence. This suggests that small bundles entirely formed of
semiconducting tubes can be ideal for opto-electronics, such as in
light-emitting devices\cite{Misewich,Chen}.

Two-photon excitation is used to derive exciton binding
energies\cite{Wang1,Maultzsch}. Fig.4 summarizes the two-photon map
of Ref.\onlinecite{Maultzsch}. The open circles are two-photon
exciton resonances. These are slightly shifted with respect to those
of Ref.\onlinecite{Wang1} due to the presence of small
bundles\cite{Maultzsch}. Below each two-photon band,
Ref.\cite{Maultzsch} reported a set of peaks, indicated by solid
squares and circles in Fig.4. These appear like a Rydberg series of
states, each matching the excitation energy of a larger gap tube, as
indicated by horizonal dashed lines in Fig.~4. These are analogous
to the energy transfer-induced peaks along each horizonal
dashed-dotted line in Fig.~2. We attribute them to emission of small
gap tubes due to exciton energy transfer from larger gap tubes in
bundles. We assign the four features in Fig.~4 with $\sim$1390 nm
excitation (solid squares) to EET from (5,4) donors to (6, 4), (9,
1), (8, 3) and (6, 5) acceptors. Since two-photon luminescence
increases quadratically with excitation power, the energy transfer
features show more distinct peaks compared to Fig.1.

\begin{figure}
\centerline{\includegraphics[width=55mm]{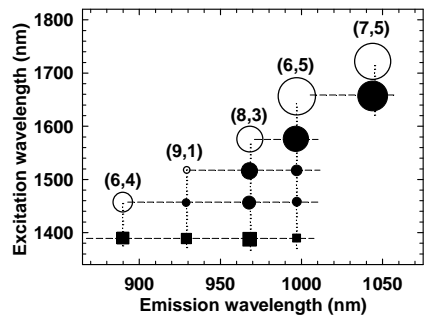}}
\caption{Some features in the two-photon map of
Ref.~\onlinecite{Maultzsch}. Open circles: two-photon peaks. Solid
circles, squares: EET peaks.}\label{fig:4}
\end{figure}

In low-dimensional systems, exciton tunneling, photon-exchange and
F$\ddot{o}$rster Resonance Energy Transfer (FRET) are efficient
exciton energy transfer mechanisms
\cite{Alexander,Forster,Kagan,Biju,Becker,Rizzo}. We attribute
energy transfer in bundles to FRET. Indeed, tunneling requires
coupling of exciton wavefunctions. Its rate decays rapidly with
donor-acceptor distance ($R_{DA}$) and is very sensitive to the
$eh_{11}$ energy difference~\cite{Alexander}. The 16 tube species in
our experiment have diameters$\sim$0.65-1.05 nm,
$eh_{11}$$\sim$0.06-0.5 eV, and chiral angle
variation$\sim$5-26$^o$\cite{Bachilo1,Bachilo2}. Therefore, the
efficiency should strongly depend on specific donor and acceptor
couples. However, the spectrum in Fig. 3b excited at (5, 4)
$eh_{11}$, reproduces the profile of the absorption in Fig.3c above
850nm, with no (n,m) preference. This suggests that the factor
dominating exciton energy transfer in bundles is tube concentration,
not symmetry, diameter or bandgap difference, thus exciton tunneling
is not the dominant mechanism.

Photon-exchange is exciton-photon coupling with no direct
donor-acceptor interaction. It has a smaller dependence on $R_{DA}$
than FRET, thus it can become significant for much longer distances
than FRET. However, the lack of significant EET features in isolated
tube solutions~\cite{Bachilo1,Bachilo2} combined with the low
quantum efficiency~\cite{Connell} suggest that even if
photon-exchange might exist between bundles or between isolated
SWNTs, it is not dominant between adjacent tubes in a given bundle.

\begin{figure}
\centerline{\includegraphics[width=95mm]{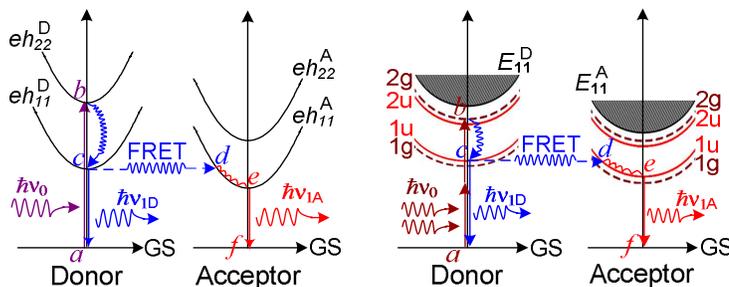}}
\caption{(Left) Schematic EET from a large gap donor (D) to a
small gap acceptor (A).($a$$\rightarrow$$b$) exciton absorption at
$eh_{22}^{D}$; ($b$$\rightarrow$$c$) fast relaxation to
$eh_{11}^{D}$ within the donor; ($c$$\rightarrow$$d$) FRET from
donor $eh_{11}^{D}$ to acceptor exciton states;
($d$$\rightarrow$$e$) fast interband relaxation down to
$eh_{11}^{A}$ of acceptor; ($e$$\rightarrow$$f$) radiative
recombination at $eh_{11}^{A}$.(Right) Recombination mechanism for
two-photon excitation  in bundles adapted from Ref.6, where 1g,
1u, 2u and 2g are the even (g) and odd (u) exciton states
associated with $E_{11}$. }\label{fig:5}
\end{figure}

FRET is a very efficient exciton energy transfer mechanism via a
resonant, near-field, dipole-dipole
interaction\cite{Forster,Kagan,Biju,Becker,Rizzo}. It is commonly
observed in biological systems, conjugated polymers, wires,
dots\cite{Forster,Kagan,Biju,Becker,Rizzo}, where it dominates at
short and intermediate distances
(0.5-10nm)~\cite{Forster,Kagan,Biju,Becker,Rizzo}. Its efficiency is
determined by the spectral overlap of donor emission and acceptor
absorption, by $R_{DA}$, and by the relative orientation of emission
and absorption dipoles~\cite{Forster}. The rate of energy transfer
is proportional to $R_{DA}^{-6}$~\cite{Forster}. The FRET efficiency
in bundles is expected to be high. Indeed, there is a large overlap
between emission of large gap tubes and absorption of small gap
tubes. SWNTs in bundles are parallel, giving a maximum dipole
orientation factor. They aggregate with small wall-to-wall
distance~\cite{Hertel,Reich3}. This makes higher multipolar
contributions possible as well~\cite{Forster,Kagan}. Indeed,
considerable luminesce quenching of CdSe-ZnS dots conjugated to
SWNTs was reported due to FRET from dots to tubes~\cite{Biju}. This
further suggests FRET to be dominant in bundles. This process is
schematized in Figs. 5a,b for both one and two-photon
spectroscopies.

In summary, we presented a thorough investigation of
photoluminescence in nanotube bundles. We have shown that the
apparently complex absorption and emission spectra can be simply
explained by exciton energy transfer between adjacent semiconducting
tubes. By studying the spectral evolution for increasing bundle
size, we assigned all the excition energy transfer peaks. We argue
that F$\ddot{o}$rster interaction between excitons dominates the
transfer process. This is highly efficient in nanotube bundles,
adding a major relaxation channel for excitons, explaining the low
luminescence yield of large gap nanotubes. Thus, contrary to what
usually assumed, bundles could be ideal for high yield
optoelectronics applications, far surpassing the poor performance of
individual tubes. Furthermore, energy transfer fingerprints bundles
with different tube concentrations, finally offering a quantitative
means to monitor the composition of solutions and films, a key
requirement for research and applications.

We acknowledge D. Prezzi, A. Rubio, A. Hartschuh for useful
discussions; funding from EPSRC GR/S97613 and Ministry of
Information and Communication, Republic of Korea (No.
A1100-0501-0073). PHT and ACF acknowledge funding from the Royal
Society. ACF from the Leverhulme Trust.

\end{document}